# Energy Efficient Routing by Switching-Off Network Interfaces

Frédéric Giroire & Dorian Mazauric & Joanna Moulierac

MASCOTTE – I3S(CNRS/University Nice Sophia Antipolis) - INRIA, France

frederic.giroire@inria.fr, dorian.mazauric@inria.fr, joanna.moulierac@inria.fr

## Abstract

Several studies exhibit that the traffic load of the routers only has a small influence on their energy consumption. Hence, the power consumption in networks is strongly related to the number of active network elements, such as interfaces, line cards, base chassis,... The goal thus is to find a routing that minimizes the (weighted) number of active network elements used when routing. In this paper, we consider a simplified architecture where a connection between two routers is represented as a link joining two network interfaces. When a connection is not used, both network interfaces can be turned off. Therefore, in order to reduce power consumption, the goal is to find the routing that minimizes the number of used links while satisfying all the demands. We first define formally the problem and we model it as an integer linear program. Then, we prove that this problem is not in APX, that is there is no polynomial-time constant-factor approximation algorithm. We propose a heuristic algorithm for this problem and we also prove some negative results about basic greedy and probabilistic algorithms. Thus we present a study on specific topologies, such as trees, grids and complete graphs, that provide bounds and results useful for real topologies. We then exhibit the gain in terms of number of network interfaces (leading to a global reduction of approximately 33 MWh for a medium-sized backbone network) for a set of existing network topologies: we see that for almost all topologies more than one third of the network interfaces can be spared for usual ranges of operation. Finally, we discuss the impact of energy efficient routing on the stretch factor and on fault tolerance.

## 1 Introduction

The minimization of ICT energy consumption has become a priority with the recent increase of energy cost and the new sensibility of the public, governments and corporations towards energy consumption. ICT alone is responsible of 2% to 10% (depending on the estimations) of the world consumption (Chiaraviglio, Leonardi, & Mellia, How much can Internet be greened?, 2009).

In this paper, we are interested in the networking part of this energy consumption, and in particular in the routing. It is estimated that switches, hubs, routers account for 6 TWh per year in the US (Nordman & Christensen, 2005).

Some recent studies (Chabarek, et al., 2008), (Mahadevan, Sharma, Banerjee, & Ranganathan, A Power Benchmarking Framework for Network Devices., 2009) exhibit that the traffic load of the routers only has a small influence on their energy consumption. Hence, the dominating factor is the number of switched-on network elements: interfaces, platforms, routers,... In order to minimize energy, we should try to use as few network elements as possible.

Nevertheless, in most of networks, PoPs or even routers cannot be turned off. As a matter of fact, first, they are the source or destination of demands; second, they can be part of backup routes to protect the network again failures. For this reason, we consider in this paper a simplified architecture where a connection between two routers is represented by a link joining two network interfaces.

We can spare energy by turning off the two network interfaces which are the extremities of the link. The network is represented by an undirected graph and, in that case, the goal in this simplified architecture is *to find a subgraph minimum in number of links to route the demands*. The contributions of this paper are the following:

- We prove that there is no polynomial-time constant-factor approximation algorithm for this problem, even for two demands or if all links have the same capacity, unless P=NP.

- We present *heuristics* to find close to optimal solutions for general networks. These heuristics are validated by comparison with theoretical bounds for specific instances of the problem. Furthermore, we exhibit negative results about greedy and probabilistic heuristic algorithms.

- We give explicit *close formulas or bounds for specific topologies*, such as trees, complete graphs, and square grids. They provide limit behaviors and give indications of how the problem behaves for general networks.

- We study the *energy gain* on a set of topologies of existing backbone networks. We exhibit that *at least one third of the network interfaces* can be spared for usual range of demands.

- Finally, we discuss the impact of energy-efficient routing on *route length* and *fault tolerance*.

## 1.1 Related Work

**Measure of energy consumptions.** Several measurement campaigns of network energy consumption have been carried out in the last few years. See for example (Mahadevan, Sharma, Banerjee, & Ranganathan, Energy Aware Network Operations, 2008) (Mahadevan, Sharma, Banerjee, & Ranganathan, A Power Benchmarking Framework for Network Devices., 2009) (Chabarek, et al., 2008). Their authors claim that the consumption of network devices is largely independent of their load. In particular, in (Chabarek, et al., 2008), the authors were interested by routers' energy consumption. They observe that for the popular Cisco 12000 series, the consumption at a load of 75% is only 2% more than at an idle state (770W vs. 755W). In (Mahadevan, Sharma, Banerjee, & Ranganathan, A Power Benchmarking Framework for Network Devices., 2009), the authors show through experimentation that the power consumed depends on the number of active ports. Explicitly disabling unused ports on a line card reduces the device power consumption. The values obtained during experimentation show that the consumption of a linecard 4-port Gigabit ethernet (100W) is approximately one fourth the consumption of the global base system (430W).

**Energy minimization.** In (Chabarek, et al., 2008), the authors model this problem as an integer linear program. The objective function is a weighted sum of the number of platforms and interfaces. They show how much energy can be saved on different networks with this model. However, they do not give intuitions, explanations, nor formulas for their results. In (Idzikowski, Orlowski, Raack, Woesner, & Wolisz, 2010), the authors propose a rerouting at different layers in IP-over-WDM networks for energy savings while (Puype, Vereecken, Colle, Pickavet, & Demeester, 2009) study the impact of the technology for energy efficient routing. In (Nedevschi, Popa, Iannaccone, Ratnasamy, & Wetherall, 2008), (Gupta & Singh, 2003), (Chiaraviglio, Leonardi, & Mellia, How much can Internet be greened?, 2009), researchers proposed techniques such as putting idle subcomponents (line cards, ports, etc.) to sleep, as well as adapting the rate for forwarding packets depending on the traffic in local area networks. In (Coudert, Nepomuceno, & Rivano, 2010), the authors propose a modulation of the radio configurations in fixed broadband wireless networks to reduce the power consumption.

We prove in this paper that there is no polynomial-time constant-factor approximation algorithm for this problem, even for specific instances. We give also theoretical results about the inefficiency of greedy and probabilistic algorithms. To the best of our knowledge, we are presenting in this paper the first study of energy-efficient routing solutions on specific topologies. We link the proposed theoretical bounds with general networks. Finally, we study the impact of such solutions on route length and fault tolerance.

The remainder of the paper is organized as follows.

- In Section 2, we first present formally the problem and we model it as an integer linear program.

- In Section 3, we recall the complexity of related problems and we prove that the problem cannot be approximated within a constant factor.

- In Section 4, we design heuristic algorithms and we prove also negative results about greedy and probabilistic heuristic algorithms.

- Then, specific topologies, such as trees or complete graphs, are discussed in Section 5,

- and grids in Section 6.

- We show the good performance of our proposed heuristic in terms of power consumption in Section 7, and we show the impact of such energy-efficient solutions on route lengths and network fault tolerance. We also propose the construction of fault-tolerant spanners.

- Finally, we discussed the impact of such algorithms for network operators in Section 8.

## 2 Problem Modeling

One way to reduce the network power consumption consists in minimizing the number of turned-on network equipments. We model a network topology as an undirected weighted graph $G=(V,E)$, where the weight $c_e$ represents the capacity of edge $e \in E$. We represent the set of demands by D={ $\mathcal{D}_{st}$>0 ;(s,t) $\in$V×V, s≠t}, where $\mathcal{D}_{st}$ denotes the amount of demand from $s$ to $t$. A demand $\mathcal{D}_{st}$ has to be routed through an elementary path from node $s \in V$ to $t \in V$. A valid routing of the demands is an assignment of such a path in $G$ for each $\mathcal{D}_{st} \in$D such that for each edge $e \in E$, the total amount of demands through $e$ does not exceed the capacity $c_e$. A classical decision problem is to determine if there is such a routing of the demands in $G$. Formally:

**Definition 2.1** *Given an undirected weighted graph G=(V,E) and a set of demands D, the* ROUTING PROBLEM *consists in deciding if there is a valid routing of the demands D in G.*

For our purpose, we study a simplified network architecture in which a link (*A*,*B*) connects two routers *A* and *B* through 2 network interfaces, one for *A* and one for *B*. The degree of a node in the graph corresponds to the number of network interfaces on the corresponding router. If a network interface is turned-off, then the other interface at the extremity of the link is not useful anymore and can also be turned-off. Therefore, the objective of our problem is to minimize the number of active links in the network. Formally:

**Definition 2.2** *Given an undirected weighted graph G=(V,E) and a set of demands D, the* MINIMUM EDGES ROUTING PROBLEM *consists in finding a minimum cardinality subset E*$\subseteq$ E such that there is a valid routing of the demands D in G.*

In Section 2.1, we present some simple instances of the MINIMUM EDGES ROUTING PROBLEM and the corresponding solutions. In Section 2.2, we describe a linear program for our problem.

## 2.1 Examples

Consider the graph *G*=(*V*,*E*) depicted in Figure 1 composed of 14 nodes and 15 edges. Integers on edges represent the different capacities. We denote by $D_1$, $D_2$, and $D_3$ the three demands $D_{s_1 t_1}$, $D_{s_2 t_2}$, and $D_{s_3 t_3}$, respectively.

In Figure 1, there are two demands $D_1 = 10$ and $D_2 = 10$. Because of the second demand, there is no feasible routing. Indeed node $s_2$ has two adjacent edges of capacities 8 and 9, and so it is impossible to route a demand of 10 units through one of these edges. In our problem, a demand can not be divided into sub-demands. The solution of the ROUTING PROBLEM for this instance is 'no', and so there is no solution for the MINIMUM EDGES ROUTING PROBLEM because the whole graph is not sufficient.

In Figure 2 (a), there are two demands $D_1 = 10$ and $D_2 = 5$. The solution $E^*$ for the MINIMUM EDGES ROUTING PROBLEM is composed of |$E^*$|=7 edges. Note that the path from $s_i$ to $t_i$ in G*=(V,E*) is composed of 5 edges, whereas the shortest path in *G* is composed of 4 edges, for $i \in \{1,2\}$. Indeed these two shortest paths are edge-disjoint whereas, in the optimal routing, the two paths share 3 edges. This simple example shows that the shortest paths routing does not give the optimal solution in general.

In Figure 2 (b), there are also two demands but the amount of demand from $s_1$ to $t_1$ is now equal to 12. This increase considerably changes the optimal solution $E^*$. Indeed the demand $D_{s_1 t_1}$ must be routed through the shortest path composed of 4 edges because of the edge of capacity 11. Thus $E^*$ is composed of the two previous edge-disjoint shortest paths of length 4. We get |$E^*$|=8.

In Figure 3, we have $D_1 = 10$, $D_2 = 5$, and $D_3 = 2$. Because of the three edges of capacity 16, only two demands can share these 3 edges. Two optimal solutions are depicted in these figures, each one of cost $|E^*|=9$. In these optimal solutions, the 3 edges of capacity 16 support the demand $D_3$ and one of the two demands $D_2$ (for Figure 3 (a)) or $D_1$ (for Figure 3 (b)). The other demand is routed through the shortest path.

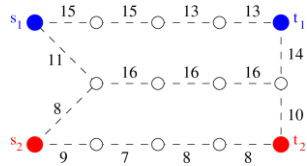

Figure 1: no solution for the MINIMUM EDGES ROUTING PROBLEM with $D_{s_1 t_1} = 10$ and $D_{s_2 t_2} = 10$.

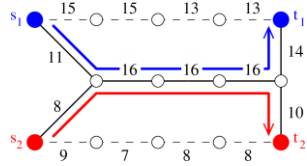

(a) $D_{s_1 t_1} = 10$, $D_{s_2 t_2} = 5$ → the two routes of the optimal solution are not shortest paths.

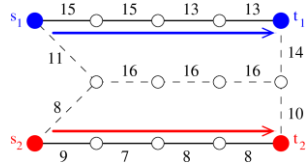

(b) $D_{s_1 t_1} = 12$, $D_{s_2 t_2} = 5$ → the optimal solution corresponds to shortest paths.

Figure 2: two different solutions for the MINIMUM EDGES ROUTING PROBLEM with almost the same set of demands.

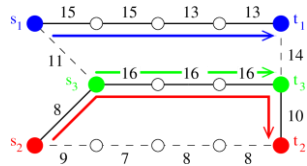

(a) $D_{s_1 t_1}$ and $D_{s_3 t_3}$ are routed through shortest paths.

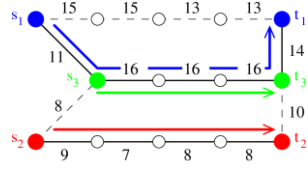

(b) $D_{s_2t_2}$ and $D_{s_3t_3}$ are routed through shortest paths.

Figure 3: Two different optimal solutions $D_{s_1t_1} = 10$, $D_{s_2t_2} = 5$, and $D_{s_3t_3} = 2$.

## 2.2 Integer Linear Program

The MINIMUM EDGES ROUTING PROBLEM can be modeled as a multicommodity integral flow problem in which the objective function is the minimization of the number of edges. We note $f^{st}_{uv}$ the flow on edge *uv* corresponding to the demand $\mathcal{D}_{st}$ flowing from *u* to *v*. We note $f_{uv} = \sum_{st \in V \times V} f^{st}_{uv}$. For each edge $e \in E$, we introduce a binary variable $x_e$ which says if the edge *e* is used or not: $x_e=0$ iff $f_{uv}+f_{vu}=0$ and $x_e=1$ iff $f_{uv}+f_{vu}>0$.

The *Objective function* is

$$\min \sum_{e \in E} x_e$$

subject to:

Flow constraints: $\forall (s,t) \in V \times V, \forall u \in V$,

$$\sum_{v \in N(u)} f^{st}_{vu} - \sum_{v \in N(u)} f^{st}_{uv} = \begin{cases} -D_{st} & \text{if} \quad u = s \\ D_{st} & \text{if} \quad u = t \\ 0 & \text{otherwise} \end{cases}.$$

Capacity constraints: $\forall e = (u,v) \in E$,

$$\sum_{d \in D} \left( f^d_{uv} + f^d_{vu} \right) \leq x_e c_e.$$

The flow constraints are usual flow conservation. The capacity constraints state that for each edge $e \in E$, the total amount of demands through *e* does not exceed the capacity $c_e$.

Table 1 summarizes the notations used throughout the paper.

| $G = (V, E)$ | Network topology with V the set of vertices (or routers) and E the set of edges (or links). |
|---|---|
| $D_{st}$ | Volume of traffic of the demand from a source $s \in V$ to a destination $t \in V$. In Section 5, $\forall (s,t) \in V \times V, D_{st} = \kappa$. |
| $c_e$ | Capacity of the edge $e \in E$. In Section 5, $\forall e \in E, c_e = c$. |
| $\lambda$ | Capacity/Demand ratio with $c = \lambda \kappa$. |
| $r_e$ | Residual capacity of edge e. |
| $OF$ | Network Overprovisionning Factor. $OF = 1$ means that the capacities $c_e$ of the edges imply the feasibility of the routing of the demands. |
| $f_{uv}^{st}$ | Flow on edge $e = uv$ corresponding to the demand $D_{st}$. |
| $x_e$ | Binary variable which says if the edge e is used or not. |
| $\bar{l}_H(D)$ | Average path length in the graph H given by a feasible routing of the demands in D. |
| $DP_H(D)$ | Average disjoint paths in the graph H for the demands in D. |
| $\delta$ | Degree of a vertex and $\delta_{max}$ is the maximum degree of the nodes in the graph. |
| $d_G(i, j)$ | Shortest path distance between node i and j in graph G. The notation $d(i, j)$ can be used for shortcut. |

Table 1: Summary of Notations

## 3 Impossibility of Approximation

The MINIMUM EDGES ROUTING PROBLEM in this paper is a special case of different well known network optimization problems: minimum cost routing (Yaged Jr.), minimum concave flow problem (Guisewite & Pardalos, 1990), minimum flow problem with step cost functions (Gabrel, Knippel, & Minoux). In operation research, this problem can be seen as a special case of the FIXED CHARGE TRANSPORTATION PROBLEM (Spielberg, 1964), (Dantzig), where the cost of the flow unit on an edge is zero. Note that this problem is NP-Hard (Guisewite & Pardalos, 1990): the number of possible subgraphs to test is strongly exponential for most graphs. Moreover, even for a given subgraph (when the set of edges to be used is fixed), the feasibility of a multicommodity integral flow problem has to be assessed. This simpler problem corresponds to the ROUTING PROBLEMit is known to be NP-complete even for two commodities (Even, Itai, & Shamir, 1975). Last, note that it is also the worst case of step-functions as most of the approximations by linearization are very far from a feasible solution.

We now prove that the MINIMUM EDGES ROUTING PROBLEM is not in APX (and so it is an NP-hard problem) even for two special kinds of instances (Theorem and Theorem ). It means that there is no polynomial-time constant-factor approximation algorithms for the MINIMUM EDGES ROUTING PROBLEM, unless *P=NP*.

**Theorem 3.1** *The* MINIMUM EDGES ROUTING PROBLEM *is not in APX even for two commodities.*

**Proof.**

Let $G=(V,E)$ be an undirected weighted graph. Consider the case of two commodities $D_{s_1t_1}>0$ and $D_{s_2t_2}>0$. We build a graph $G'=(V',E')$ as follows: we start with a copy of $G$; we add a path $P_1$ between $s_1$ and $t_1$ composed of $x>k|E|$ edges and a path $P_2$ between $s_2$ and $t_2$ also composed of $x>k|E|$ edges. Each edge of $P_1$ and $P_2$ has an infinite capacity. We consider the same set of demands. $G'$ is depicted in Figure 4. Suppose now that there is a constant $k \geq 1$, such that there exists a polynomial-time algorithm finding a subset $E^k \subseteq E$ ensuring that $|E^k|/|E^*| \leq k$, where $E^*$ is an optimal solution of the MINIMUM EDGES ROUTING PROBLEM. We name $E'^k$ the solution found for $G'$ by this algorithm and we have $|E'^k|/|E'^*| \leq k$. If $|E'^k| \geq x$, then it means that there is no solution for the ROUTING PROBLEM $G$. As a matter of fact, such solution would use at most the $|E|$ edges of $G$, and a $k$-approximation would use less than $x$ edges. If $|E'^k|<x$, then it means that there is a solution for this problem. Thus, using this construction, we get a polynomial-time algorithm solving the ROUTING PROBLEM $G$, that is a polynomial-time algorithm to decide if there is a routing of the demands $D$ in $G$ respecting the constraints of capacities. A contradiction, unless $P=NP$.

□

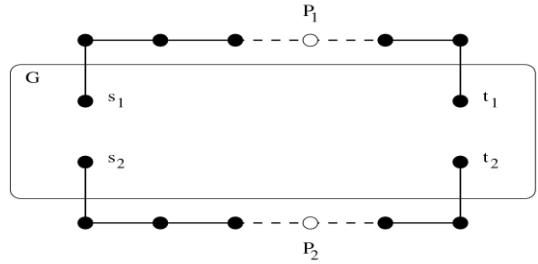

Figure 4: Construction of $G'$ from $G$ used in the proof of Theorem 3.1.

**Theorem 3.2** *The MINIMUM EDGES ROUTING PROBLEM is not in APX even if each edge has a constant capacity c.*

**Proof**

Suppose that there is a constant $k \geq 1$, such that there exists a polynomial-time algorithm finding a subset $E^k \subseteq E$ for the MINIMUM EDGES ROUTING PROBLEM ensuring that $|E^k|/|E^*| \leq k$. Recall that $E^*$ denotes an optimal solution of the MINIMUM EDGES ROUTING PROBLEM. Consider the undirected weighted graph $G=(V,E)$ depicted in Figure 5. Each edge $e \in E$ has capacity $c_e=c$. The $|D|$ demands are such that $\sum_{i=1}^{|D|} D_{s_it_i} = 3c$. From $u$ to $v$, there are 3 disjoint paths composed of 2 edges each and $|D|$ disjoint paths of $x > k|E|$ edges each. Because of the $k$-approximation, if there is a solution using no long paths, then our polynomial-time algorithm returns necessarily such a solution. Otherwise it returns a solution with edges belonging to these paths. Note that the problem of finding if there exists a partition of requests into three sets of same weight $c$ is an NP-complete problem. Our polynomial-time approximation algorithm would thus solve it, a contradiction, unless $P=NP$.

□

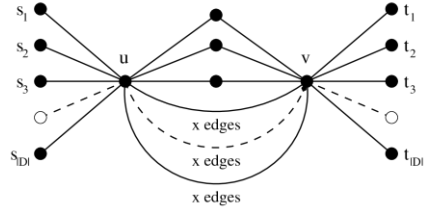

Figure 5: Graph $G=(V,E)$ used in the proof of Theorem 3.2. Each edge $e \in E$ has capacity $c_e=c$.

These two negative results motivate the design of heuristic algorithms for the MINIMUM EDGES ROUTING PROBLEM in Section 4 and the study of theoretical bounds for particular instances in Section 5 that give information for general networks.

## 4 Heuristics

As we have seen in the previous section, the MINIMUM EDGES ROUTING PROBLEM is a problem difficult to solve exactly and is even difficult to approximate to an insured factor in general. Hence, the necessity of proposing good heuristic algorithms for classical real network topologies. We propose in this section two heuristics to find energy-efficient routing, namely LESS LOADED EDGE HEURISTIC, RANDOM HEURISTIC. These heuristics are tested in Section 7. We also discuss the efficiency of greedy and random heuristics on the MINIMUM EDGES ROUTING PROBLEM.

### 4.1 Heuristics

Algorithm 1 presents a simple heuristic named LESS LOADED EDGE HEURISTIC for our problem. We start from the whole network, compute a feasible routing as described in Algorithm 2 and try to remove in priority edges that are less loaded. We believe that it is better to remove these edges that are not involved in many shortest paths than overloaded edges. For the routing, the demands are considered one by one in random order. We compute a shortest path for the demand with the metric $c_e/r_e$ on edges, where $r_e$ is the residual capacity on edge $e$ when the previous demands have been routed. Then, the residual capacity is updated for each edge and the next demand is considered. This routing allows a better load balancing of the demands in the network. Note that finding a feasible routing can also be done with an integer linear program for the ROUTING PROBLEM for small topologies. Each time an edge is removed in the network, a feasible routing is computed. If no routing exists, then the removed edge is put back and we try to remove another edge that has not been yet considered. The process of removing less loaded edges is done until no more edges can be removed.

The second heuristic, RANDOM HEURISTIC, is used as a measure of comparison during the simulations. The only difference with the first one is that it selects uniformly at random the links to be removed and not (necessarily) the less loaded edges. The routing is performed in the same way as for LESS LOADED EDGE HEURISTIC.

**Algorithm 1** LESS LOADED EDGE HEURISTIC

**Require:** An undirected weighted graph $G = (V, E)$ where each edge $e \in E$ has an initial capacity $c_e$ and a residual capacity $r_e$ (depending on the demands supported on e). A set of demands $D$, each demand has a volume of traffic $D_{st}$.

$\forall e \in E, r_e = c_e$
Compute a feasible routing of the demands with Algorithm 2
**while** Edges can be removed **do**
Remove the edge e' that has not been chosen once, with the smallest $c_{e'}/r_{e'}$.
Compute a feasible routing with Algorithm 2
If no feasible routing exists, then put back e' in $G$
**end while**
**return** the subgraph $G$.

---

**Algorithm 2** Feasible routing Heuristic for ROUTING PROBLEM

**Require:** An undirected weighted graph $G = (V, E)$ where each edge $e \in E$ has an initial capacity $c_e$ and a residual capacity $r_e$ (depending on the demands supported on e). A set of demands $D$, each demand has a volume of traffic $D_{st}$.

Sort the demands in random order
**while** $D_{st}$ is a demand in $D$ with no routing assigned **do**
Compute a shortest path $SP_{st}$ with the metric $c_e/r_e$ on edges
Assign the routing $SP_{st}$ to the demand $D_{st}$
$\forall e \in SP_{st}, r_e = c_e - D_{st}$
**end while**
**return** the routing (if it exists) assigned to the demands in $D$.

---

These heuristics are evaluated through simulations in Section 6.3 and 7 and compared to the theoretical bounds given in Section 5 and to the integer linear program described in Section 2.2.

## 4.2 Note on the execution time of Random Heuristic

Consider an undirected weighted graph $G=(V,E)$, a set of demands $D$, and a unique optimal solution $E^*$ for the MINIMUM EDGES ROUTING PROBLEM.

When you have drawn a percentage $\mu$ of edges of $G$ and when the routing is still feasible after these edges removals, the probability of having removed an edge of the optimal solution is given by

$$1 - \frac{\binom{|E| - |E^*|}{\mu|E|}}{\binom{|E|}{\mu|E|}}.$$

For example, if |E|=100 and |E*|=20, then the probability to remove 50 edges without removing one edge of the solution is smaller than 8.8 $10^{-8}$. The number of random trials to find the optimal solution is then prohibitive for large networks.

**Order of the execution time.** A bound can be easily derived by considering the drawing problem but with replacement. The probability to draw an edge of the solution is larger than

$$1-\left(1-\frac{|E^*|}{|E|}\right)^{\mu|E|}.$$

When |E| is large, we have

$$\approx 1-e^{-\mu|E^*|}.$$

Hence, finding the optimal solution by removing edges uniformly at random may be exponentially long.

When the number of vertices is large, the time of execution of the random heuristic may be very large. Indeed, $\exp(-.5\times50)=1.4\ 10^{-11}$. Tens of billions of trials would be necessary to find the optimal solution for a graph with only one optimal solution.

## 4.3 Shortest Paths Routings and Greedy Algorithms

In this section we show that greedy algorithms based on shortest paths or minimum added edges at each step, may return a solution whose cardinality is arbitrarily large compared to the optimal number |E*|. Furthermore, even if we have the best order that minimizes the total number of edges used when routing, the quality of the solution may be very bad compared to the optimal one.

**Definition 4.3.1** *Given an undirected weighted graph G=(V,E) and a set of demands D, $E^{sp} \subseteq E$ is a minimum cardinality set such that there is a valid routing of the demands D in $G^{SP}=(V,E^{SP})$ and such that each demand is routed through a shortest path of G.*

Lemma proves that $E^{SP}$ may give a number of edges $|E^{SP}|$ arbitrarily large compared to the number of edges |E*| of an optimal solution E* of the MINIMUM EDGES ROUTING PROBLEM.

**Lemma 4.3.2** *For any C>1, there is an instance such that $|E^{SP}|>C|E^*|$.*

**Proof**

Consider the undirected weighted graph G=(V,E) depicted in Figure 6. Integers on edges represent the different capacities. The |D| demands are each of value 1. $\forall i \in [1, D]$, the length of a shortest path from $s_i$ to $t_i$ is 5. If each demand is routed through a shortest path, then it is easy to see that the routing uses $|E^{SP}|=5|D|$ edges because all the edges of all the shortest paths have capacity 1. Indeed any two demands are routed through edge-disjoint shortest paths. However, there is an optimal routing $E^*$ for the MINIMUM EDGES ROUTING PROBLEM such that $|E^*|=2|D|+4$: each demand is routed through the 4 edges of infinite capacity. Note that for this optimal routing, there is no demand routed through a shortest path. Taking $|D|>4C/(5-2C)$, we get the inequality.

Furthermore if the number $|D|$ of demands is fixed, we can modify $G$ as follows: we replace the path of 4 edges from $u$ to $v$ by a path of $x+1$ edges, each of infinite capacity; we replace the $|D|$ disjoint paths of 3 edges each from $u$ to $v$ by $|D|$ disjoint paths of $x$ edges, each of capacity 1. We get $|E^{SP}|=|D|(2+x)$ and $|E^*|=2|D|+x+1$. Taking $x > (2|D|(C-1)+1)/(|D|-C)$, we get the inequality.

□

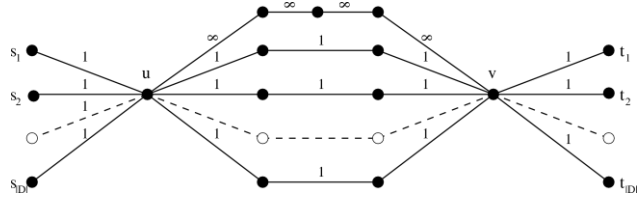

Figure 6: Undirected weighted graph used in Lemma 4.3.2. From $u$ to $v$, there are $|D|+1$ elementary paths: 1 is composed of 4 edges of infinite capacity and $|D|$ are composed of 3 edges of capacity 1. There are $|D|$ demands, each of value 1.

Consider the instance of the MINIMUM EDGES ROUTING PROBLEM described in proof of Lemma composed of $|D|$ demands. Consider any order of these demands among the $|D|!$. From Lemma , we deduce that if we route the demands greedily according to this order, through a shortest path, then we get a routing that uses $5|D|$ whereas $|E^*|=2|D|+4$. In other words, even if we have the best order to route greedily the demands, the number of edges may be arbitrarily large compared to the minimum number. Furthermore, if we route the demands greedily choosing, at each step, a path that minimizes the number of added edges, then we get also $5|D|$ edges for any order.

In conclusion simple greedy algorithms may return a solution whose cardinality is arbitrarily large compared to the optimal number $|E^*|$. Even if we have the best order, the quality of the solution may be very bad compared to the optimal one.

Because of previous negative results, we study in Section 5 the MINIMUM EDGES ROUTING PROBLEM for specific instances.

## 5 Topology Study: extreme cases

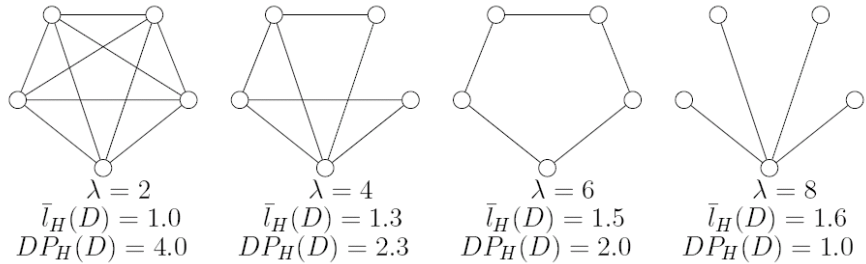

Figure 7: A Toy example, study of the complete graph with 5 vertices: subgraphs with the minimum number of edges, with λ the capacity/demand ratio, $\overline{l}_H(D)$ average route length and $DP_H(D)$ the average number of edge-disjoint paths between two nodes where $D$ contains the all-to-all demands.

We present here the general framework of the studies of the rest of this paper. We then study two extreme cases for general networks, namely trees and complete graphs. They give us the limit behavior of the real networks for different network loads. We also derive an upper bound on the number of links that can be spared in function of the demands.

## 5.1 Our Framework

In general networks, the demands vary during the life of the networks, e.g. with the increase of the number of users, with the development of new technologies, or, more simply, according to the time of the day. The goal of the study in this section is to see how much energy can be spared depending on the resources available for the routing.

Given an undirected weighted graph $G=(V,E)$ representing a network, each edge $e \in E$ has the same amount of capacity $c_e = c$. We perform an all-to-all routing with demands of volume, $\forall (s,t) \in V \times V$, $\mathcal{D}_{st} = \kappa$.

**Definition 5.1.1** *The capacity/demand ratio λ expresses the relation between the demand and the edge capacity: $\lambda = c/\kappa$.*

Although an all-to-all routing is not a realistic scenario, it allows to study the effects of such routing in extreme conditions and also to keep the network connected. With such scenario, we can certify that the topology given by our algorithm will be suitable for more realistic traffic matrices. Then, for each topology, we look at the number of links that can be spared for different capacity/demand ratios λ.

**A toy example.** As an illustration, Figure 7 shows an optimal solution $H$ given by the integer linear program described in Section 2.2 for the complete graph $G$ composed of 5 nodes when the capacity/demand ratio λ varies from 2 to 8. When λ equals 2, all the edges of $G$ are needed to perform an all-to-all routing with κ=1. The larger the ratio is, the fewer network interfaces are needed until we reach the star graph where no more network interfaces can be removed. The two extreme cases are the complete graph (λ=2) and the tree (λ=8). The gain in terms of network interfaces can reach 60%, indeed, only 4 edges (or 8 network interfaces) are needed instead of 10 (or 20). We also measure the impact of this energy-efficient routing on delay and failure protection: for each different λ, we give the *average route length*, $\overline{l}_H(D)$, given by a feasible routing of the all-to-all demands on the solution subgraph $H$, and the *average number of edge-disjoint paths* linking two nodes, $DP_H(D)$. We see that, for this simple example, the route length increases by 60% and that the number of disjoint paths drops from 4 to 1 between the two extreme cases.

Note that an orthogonal way to conduct the study is to fix the number of network interfaces/edges to be turned-on and to compute the *load of the network*, that is the minimum capacity needed to be able to satisfy the all-to-all routing.

**Definition 5.1.2 (Load of a Graph)** *Let $G=(V,E)$ be an undirected weighted graph and D the set of demands. The load of G is the minimum over all routings (feasible flows $\mathcal{F}$) of the maximum load over all edges:*

$$\min_{f \in F} \max_{e \in E} f_e.$$

## 5.2 General Bounds

**Path length lower bounds.** The global capacity of the system has to be larger than the global demand. The global flow is minimum when all demands are following the shortest paths between their source and destination. We note $d(s,t)$ the length of a shortest path between a source $s$ and a destination $t$. Hence, we have

$$\sum_{e \in E} c_e \geq \sum_{st \in V^2; s \neq t} d(s,t) D_{st}.$$

In particular, when all the edges have the same capacity $c$ and all the couples of nodes have the same demand κ, it becomes

$$c|E| \geq \kappa \sum_{st \in V^2; s \neq t} d(s,t).$$

or equivalently

$$|E| \geq \frac{1}{\lambda} \sum_{st \in V^2; s \neq t} d(s,t).$$

**Max flow min cut lower bounds.** We present here a generalized max flow min cut argument. For each subset $S \subseteq V$, we must have

$$\sum_{e=uv \in E; u \in S, v \in \overline{S}} c_e \geq \sum_{s \in S, t \in \overline{S}} D_{st} + \sum_{s \in S, t \in \overline{S}} D_{ts}.$$

where $\overline{S} = V - S$. In particular, when all the edges have the same capacity $c$ and all the couples of nodes have the same demand κ, it becomes

$$c|E_{S\overline{S}}| \geq 2\kappa |S||\overline{S}|,$$

where $|E_{S\overline{S}}|$ is the number of edges of the cut between $S$ and $\overline{S}$. The load of a graph can thus be computed by looking at the *minimum cut* of the network that supports the maximum flow.

*Minimum bisection cut.* A particular example of cut is the minimum *bisection cut* denoted $|E_{S\overline{S}}|$. The minimum bisection cut is the minimum cut that divides the network into two (almost) equal-sized regions $S$ and $\overline{S}$:

$$|S| = \left\lceil \frac{n}{2} \right\rceil \text{ and } |\overline{S}| = \left\lfloor \frac{n}{2} \right\rfloor.$$

The set of edges corresponding to the minimum bisection cut will support the demands exchanged between the nodes of the two regions. This gives a minimum value for the load of the graph which is for most practical networks a good approximation of the real load as shown in Section 7.

In a graph with $n$ vertices, with minimum bisection cut $|E'_{S\overline{S}}|$, the load of the graph in case of all-to-all demand is at least:

$$\frac{2\kappa}{|E'_{s\bar{s}}|}\left\lceil\frac{n}{2}\right\rceil\left\lfloor\frac{n}{2}\right\rfloor.$$

## 5.3 Load of the Minimal Subgraph: a Spanning Tree

In the undirected case, the subgraph with the minimum number of edges is a tree, as it is the smallest connected subgraph, see e.g. Figure 7. This minimal configuration can be attained when the capacity is larger than the load given in Lemma 5.3.1.

**Lemma 5.3.1** *[Tree and Spanning Tree]*

> *a) The load of a tree composed of n nodes is $2(n-v)$, where v is the size of the larger branch incident to the tree centroid.*
>
> *b) In a graph with n nodes and of maximum degree $\delta_{max}$, the load of a spanning tree is at least*
>
> $$2\kappa\left\lceil\frac{n-1}{\delta_{max}}\right\rceil\left(n-\left\lceil\frac{n-1}{\delta_{max}}\right\rceil\right).$$

**Proof**

**a)** *Load of an edge.* Removing an edge $e$ in a tree disconnects the graph into 2 connected components of sizes $v(e)$ and $n-v(e)$ (with $v(e) \leq n-v(e)$). The load of this edge is $2(e)(n-v(e))$.

*Tree centroid.* Consider the tree centroid $C$ and consider one of the branch starting at $C$. A tree centroid is a vertex or an edge $C$ which minimizes over all nodes the largest connected component induced by removing $C$ from the graph. The maximal load of an edge of this branch is the load of the edge $e^*$ incident to $C$. Indeed, for any $e$ of the branch $v(e) < v(e^*)$ and $v(e^*) \leq \lfloor(n+1)/2\rfloor$, as any branch of the centroid is of size less than $\lfloor(n+1)/2\rfloor$. Hence $2v(e)(n-v(e)) \leq 2v(e^*)(n-v(e^*))$, as the function $x(n-x)$ is increasing for $0 \leq x \leq n/2$. For the same reasons, the load of the tree is the load of the edge of the centroid (if the centroid is an edge) or incident to the centroid in the larger branch.

**b)** We have just shown that the load of a tree is the load of the edge centroid or incident to the centroid in the larger branch. The load is then minimum for a tree with minimum larger branch. In a graph of maximum degree $\delta_{max}$, the larger branch is of size at least $\lceil(n-1)/\delta_{max}\rceil$. Hence, the load of a spanning tree is at least

$$2\kappa\left\lceil\frac{n-1}{\delta_{max}}\right\rceil\left(n-\left\lceil\frac{n-1}{\delta_{max}}\right\rceil\right).$$

□

Note that the load of a (spanning) tree mostly depends on the maximum node degree of the underlying graph. On a complete graph where $\delta_{max} = n-1$, the tree with the lowest load is a star and its load is $2\kappa(n-1)$, to be compared with the load of a path $\kappa \lceil n^2/2 \rceil$, for which $\delta_{max}=2$. Hence, networks with nodes of large degrees tend to attain the minimum configuration for smallest capacities, see Section 7.

## 5.4 Complete Graph - Bound on the Number of Spared Edges

We consider here the complete graph $K_n$ composed of $n$ nodes and $n(n-1)/2$ edges. This topology is the *second extreme case* (all possible edges) of our problem. It corresponds to the *design problem* of finding the best network satisfying the load when all possible edges between nodes can be used.

The all-to-all routing is possible on the complete graph as soon as the capacity/demand ratio $\lambda$ is larger than 2. We also have seen in the previous section that when $\lambda$ is larger than $2(n-1)$ the routing is possible on the star with only $n-1$ edges, corresponding to the fraction $2/n$ of the total number of edges of the complete graph. But what happens between these two capacities?

**Lemma 5.4.1** *In a complete graph with n vertices, the all-to-all routing uses at least*

$$\max\left(\frac{2\kappa n(n-1)}{c+2\kappa}, n-1\right)$$

*edges, with c the capacity of the edges.*

**Proof**

When all the edges have the same capacity $c$ and for all-to-all demands,

$$|E^*|c \geq \kappa \sum_{i,j \in V \times V} d(i,j),$$

where $d(i,j)$ is the shortest path distance between $i$ and $j$. Note that $\sum_{i,j \in V \times V} d(i,j)$ is given as a closed formula by twice the Wiener Index (Wiener, 1947).

In a complete graph, all the paths have length one. If we remove the edge $ij$, we know that two paths (from $i$ to $j$ and from $j$ to $i$) are now of length at least 2. Hence, for an optimal subgraph with $|E^*|$ edges, we have at most $2|E^*|$ paths of length 1, and $2(n(n-1)-2|E^*|)$ paths of length at least two. This gives

$$|E^*|c \geq \kappa \sum_{ij \in V \times V} d(i,j) \geq 2\kappa|E^*| + 2\kappa(n(n-1) - 2|E^*|).$$

The bound, holding for any topology, gives

$$|E^*| \geq \frac{2\kappa n(n-1)}{c+2\kappa}.$$

□

We validate in Figure 8 the results of the integer linear program described in Section 2.2 given by CPLEX 10 for the complete graph with 5 nodes. The figure shows that the lower bound given in Lemma 5.4.1 is close to the optimal solution. The capacity/demand ratio $\lambda$ varies between 2 and 8 as stated before. For $\lambda=4$, a gain of 30% of networks interfaces is attained, leading to 7 active links instead of 10.

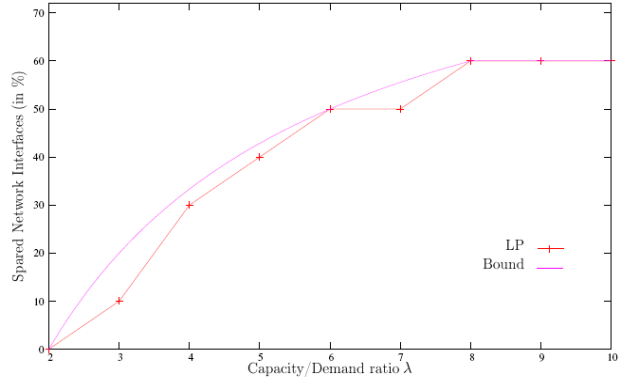

Figure 8: Number of spared edges on the complete graph with 5 nodes. Bound of Lemma 5.4.1 and integer linear program described in Section 2.2.

## 6 Results on the Square Grid

After providing some bounds on extreme cases, such as trees or complete graphs, we consider in this section MINIMUM EDGES ROUTING PROBLEM on the square grid which is an example of a structured network studied, e.g., in the context of wireless mesh networks (Akyildiz, Wang, & W., 2005). We give the load of the best spanning tree and of the full grid when all the edges are present in Section 6.1. We then propose bounds and constructions for intermediate loads in Section 6.2. Last, we compare the proposed constructions with the solutions given by the integer linear program and the heuristics LESS LOADED EDGE HEURISTIC and RANDOM HEURISTIC.

## 6.1 Limit Configurations

We first look at the load of the two limit configurations with minimum and maximum number of edges, namely spanning trees and full grids. Note that in the following, for sake of simplicity, results are presented for κ=1.

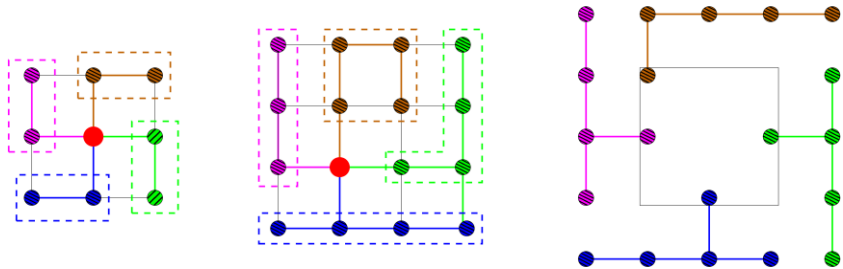

Figure 9: Partition of the grid into 4 connected subsets of almost equal sizes. Left: 3×3 grid, Middle: 4×4 grid, Right: Induction step of Lemma 6.1.1.1.

### 6.1.1 Spanning Tree

The load of the best (accepting the largest volume of demands) spanning tree is given by the following lemma.

**Lemma 6.1.1.1 (Grid Spanning Tree)** *The tree with the smallest load on a $a \times a$ square grid ($a \geq 3$), is a tree with a centroid of degree 4 and 4 branches of almost equal sizes. Its load is*

$$2 \left\lceil \frac{a^2-1}{4} \right\rceil \left( a^2 - \left\lceil \frac{a^2-1}{4} \right\rceil \right) \approx \frac{3a^4}{8}.$$

**Proof**

In any $a \times a$ square grid, it is possible to partition the vertices into a central node $C$ and 4 connected subsets of sizes $\lceil (a^2-1)/4 \rceil$ and $\lfloor (a^2-1)/4 \rfloor$. The proof is done by induction. The induction hypothesis $H_a$ is: The partition $P=\{P_1, P_2, P_3, P_4\}$ exists for an $a \times a$ grid and part $i$ of the partition is adjacent to side $i$ of the square. It is true for the $3 \times 3$ grid and the $4 \times 4$ grid, see Figure 9. We prove that $H_a$ implies $H_{a+2}$ by connecting the side $i$ of the square to one of the vertices of part $P_i$ that is on side $i$.

Hence, there exists a spanning tree of centroid $C$, with four branch and whose largest branch is of size $\lceil (a^2-1)/4 \rceil$. Its load is then $2\lceil (a^2-1)/4 \rceil (a^2-\lceil (a^2-1)/4 \rceil)$ and is maximum, as stated in Lemma 5.3.1.

□

### 6.1.2 Full grid

The maximum volume of demands that can be routed on a square grid is given by a cut argument in the following lemma.

**Lemma 6.1.2.1 (Load on a $a \times a$ grid)** *The edge load on a grid is larger than, for $a$ even,*

$$\frac{a^3}{2},$$

*and, for $a$ odd,*

$$\frac{a^3}{2}\left(1 - \frac{1}{a} + \frac{1}{a^2} - \frac{1}{a^3}\right).$$

**Proof**

From the second general lower bound argument, with a cut "in the middle" of the grid. We divide the grid into two almost equal subsets.

If $a$ even, we have $a^2/2$ nodes in $S$ and in $\bar{S}$. We have $a$ edges in the cut. Thus, we have $ac \geq 2\ a^2/2\ a^2/2$, giving $c \geq a^3/2$

If $a$ odd, we have $\lfloor a^2/2 \rfloor$ nodes in $S$ and $\lfloor a^2/2 \rfloor +1$ in $\bar{S}$. But we have here $a+1$ edges in the cut. Hence,

$$(a+1)c \geq 2 \left\lfloor \frac{a^2}{2} \right\rfloor \cdot \left( \left\lfloor \frac{a^2}{2} \right\rfloor + 1 \right) = 2\frac{a^2-1}{2}\left(\frac{a^2-1}{2}+1\right) = \frac{(a^2+1)(a^2-1)}{2}.$$

It gives
$$c \geq \frac{(a+1)(a-1)(a^2+1)}{2(a+1)} = \frac{a^3}{2}\left(1 - \frac{1}{a} + \frac{1}{a^2} - \frac{1}{a^3}\right).$$
□

## 6.2 Intermediate configurations

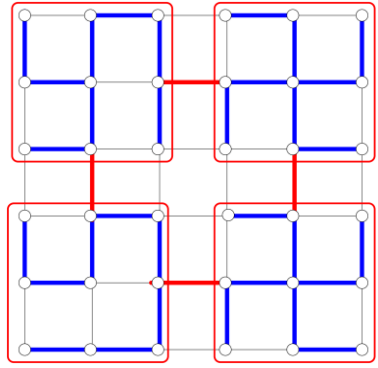

Figure 10: Construction of a subgraph with $n+p-\sqrt{p}$ edges: 1- Divide the grid into $p$ almost equal-sized regions ($p=4$ here). 2- Link the regions into a grid (red edges). 3- Connect nodes inside a region into a tree (blue edges).

We have already got the load of the two extremes cases: the full grid (when we have $2n-2\sqrt{n}$ edges) and a spanning tree (when we have $n-1$ edges). We know that the load spans between two different orders from $n^{3/2}$ to $n^2$. How the load evolves between these two values? We give a bound on the load based on the computations of the shortest paths at the end of the section. But first, we present constructions of subgraphs spanning between these two orders.

### 6.2.1 Constructions

We present here a construction of a subgraph with $n+p-2\sqrt{p}$ edges for $p$ a square number, that is if $\exists \in \mathbb{N}, p=q^2$. The load of the subgraph is of order $n^2/\sqrt{p}$. This construction gives an upper bound of the load of a grid with missing edges. For other values of $p$ ($-1 \leq p \leq n$), this formula is taken as an approximation of the load of the best subgraph. Note that it spans between the orders of the two limit configurations. We discuss this approximation for two examples, the 4×4 and the 10×10 square grids.

**Claim 6.2.1.1 (Square Grid)** *Let p be a square number between 1 and n. The load of a subgraph with n+p edges is less than*

$$\frac{1}{4}\frac{n^2}{\sqrt{p}} + \frac{3}{8}\frac{n^2}{p^2}.$$

**Proof**

The proof is done by providing a construction with this load. We divide the grid into *p* regions, see Figure 10. The regions are connected into a *square grid*. The node inside a region are linked into a *tree*. The graph has $n+p-2\sqrt{p}$ edges ($n-1-(p-1)$ edges inside the regions and $2p-2\sqrt{p}$ edges of the grid). Note that when $p=n$, the construction is simply the full square grid. Similarly, when $p=1$, the construction is just a spanning tree.

*Load of the edges of the square grid, $L_{grid}$*. The load of a grid with $p$ nodes is $\frac{1}{2}p^{3/2}$. Each region corresponds to $\lceil n/p \rceil$ or $\lfloor n/p \rfloor$ nodes. Hence, between two nodes, there are (at most) $\lceil n/p \rceil \times \lfloor n/p \rfloor$ routes instead of 1. (By choosing the routing of the grid for the $\approx n^2/p^2$ demands), we can route with a capacity of

$$\frac{1}{2}\frac{n^2}{p^2}p^{3/2}.$$

*Load of the edges inside the regions $L_{region}$*. There are at most $4L_{grid}$ entering/leaving node. These demands have to be routed through the region. The spanning tree of a region can be constructed in a way that the (at most) 4 edges of the grid are connected by 4 branches to a node of the region with degree 4. Hence, it is possible to route this demand with a capacity of $L_{grid}$.

In addition, demands have to be routed between nodes inside the region. This corresponds to an all-to-all routing in a tree of size $\lceil n/p \rceil$. An additional capacity of $\approx 3/8.n^2$ is needed.

In summary, we get

$$L_{region} \leq \frac{1}{2}\frac{n^2}{\sqrt{p}} + \frac{3}{8}n^2.$$

□

### 6.2.2 Lower Bound

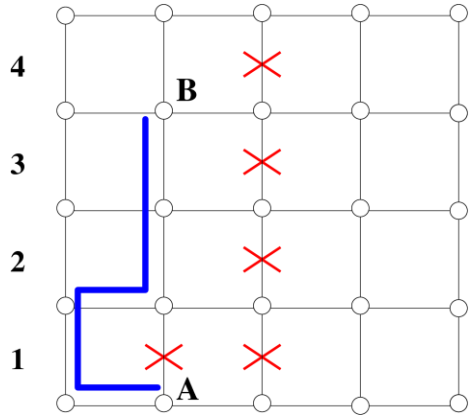

Figure 11: Proof of Lemma 6.2.2.1. A grid with 5 edges removed: the edges are removed in a column (raw) starting from position 1. The shortest path between nodes A and B increases by 2 (blue path).

We give here a bound on the load based on the computations of the shortest paths.

**Lemma 6.2.2.1** *Let G be an a×a square grid. The load of a subgraph of G with k edges is at least*

$$\frac{1}{k}\left(\frac{2a^3(a^2-1)}{3} + \left\lfloor\frac{|E|-k}{a}\right\rfloor 2a(a-1) + 4\sum_{i=1}^{|E|-k-\left\lfloor\frac{|E|-k}{a}\right\rfloor(a-1)}(a-i)\right).$$

## Proof

The sum of the shortest path lengths for a full grid is directly obtained from the Wiener index (H. Wiener. (1947) ; Eric W. Weisstein. (w.d.)) of a grid. We multiply it by two, as, in our case, we have an all-to-all demand, that is we have both path $(u,v)$ and path $(v,u)$. We get $(2a^3 \cdot (a^2-1))/3$.

When you remove an edge from the grid, all the shortest paths that were using this edge are increased by 2 units, see Figure 11. For an edge at position $i$, there are at least $2i(a-i)$ such paths on the column (or raw) of the edge. It is minimum when $i=1$ and equal to $2(a-1)$. To obtain a lower bound on the load, we have to take a sequence of edge removal that minimizes the increase of the sum of the shortest paths. It is minimum to remove edges of a column (raw) with missing edges. The second edge removal increases at least $2(a-2)$ paths, and more generally the $i$th removal increases $2(a-i)$ paths. In total, at least $a(a-1)$ paths are increased when a whole column (raw) has been removed.

In a subgraph with $k$ edges, where $|E|-k$ edges have been removed from the original graph, $\left\lfloor\frac{|E|-k}{a}\right\rfloor$ such columns (raws) can be removed. It gives the term

$$2\left\lfloor\frac{|E|-k}{a}\right\rfloor a(a-1).$$

The remaining edges are removed from the same column (raw) giving the term:

$$2\sum_{i=1}^{|E|-k-\left\lfloor\frac{|E|-k}{a}\right\rfloor(a-1)}(2a-2i).$$

□

## 6.3 Results

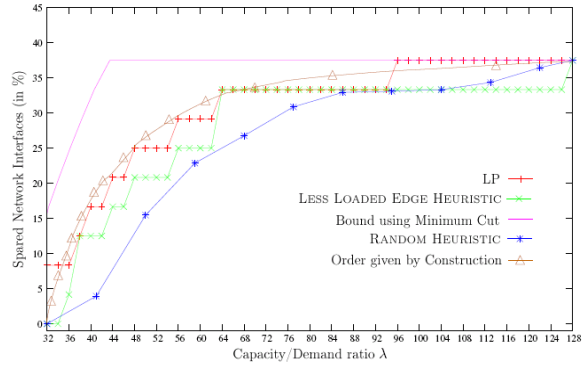

Figure 12: Percentage of spared network interfaces for a 4×4 grid.

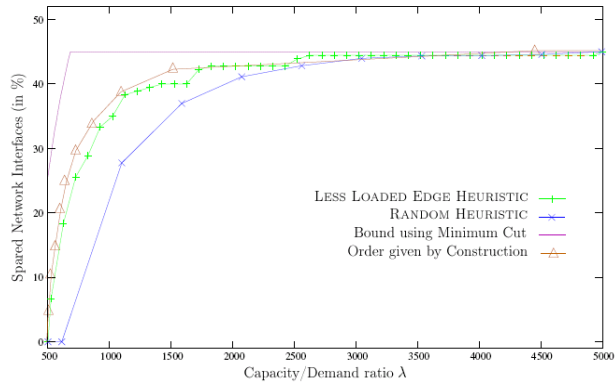

Figure 13: Percentage of spared network interfaces for a 10×10 grid.

We present results for the 4×4 grid (LP, lower bound, construction and heuristics) and for a 10×10 grid (lower bound, construction and heuristics), see Figure 12 and 13.

**Bound using Minimum Cut.** On Figure 12, we give result for the 4×4 grid. We have plotted the path length bound (magenta −). We see that, on the contrary to the complete graph, that this bound is not very tight. The explanation is that edges in the square grid are not all equivalent: the edges in the middle of the grid have a larger load than those on the borders with a shortest path routing.

The load can be computed by looking at the *minimum cut* of the network as explained in Lemma 6.1.2.1. The cut composed by the $a$ horizontal edges in the middle of the grid gives $c \geq a^3/2$. For the 4×4 grid, we can check that the load induced by the routing is greater than $4^3/2 = 32$. This cut phenomena also explains that as soon as the routing is possible (for $\lambda = 32$), some edges can be spared (around 8%): the middle edges are fully loaded, but not the edges on the borders. The same cut phenomenas are present for some SNDLib topologies studied in the next section.

**Order given by construction.** This order, given by Claim 6.2.1.1, is depicted on Figure 12 and 13. We see that it is very close (only a few percent) to the optimal solution given by the integer linear program and largely better than the bound using minimum cut. For the grid 10×10, we were not able to launch the integer linear program, and we see that LESS LOADED EDGE HEURISTIC behaves well compared to this order.

**How many network interfaces can be spared?** For values of λ larger than 32, a large number of edges can be saved at the beginning: for the 4×4 grid, 25% for an overprovisioning factor of 1.5 (λ=48) and 33% for a factor of 2 (λ=64). The savings become less important for larger capacities: only a 4 percent difference between capacities of 64 and 96, and the tree is attained with an overprovisioning factor of 3 (λ=96). For this value of λ, we use $a^2-1=15$ edges, when the full grid has 24 edges, saving 37.5% of edges.

**Behavior of the heuristics.** Note that the heuristic (green ×) behaves well and is close to the optimal values given by the integer linear program (red +). The heuristic is also significantly better than the *Random* heuristic, see Figure 12 and 13. For example, for the 10×10 grid, for a capacity/demand ratio of 1000, LESS LOADED EDGE HEURISTIC saves 39% and RANDOM HEURISTIC saves 28%.

# 7 Results on General Networks

We present in this section the results of our proposed heuristics on general networks. We study ten classical network topologies extracted from SNDLib (http://sndlib.zib.de). In our experiments, we explore how many network interfaces can be spared for different ranges of overprovisioning factor. We consider a range of capacity/demand ratio λ starting from the smallest value $\lambda_1$ allowing to route all the demands (overprovisioning factor equals 1) to the value $\lambda_{tree}$ allowing to route on a minimal subgraph, that is a spanning tree (overprovisioning factor equals $\lambda_{tree}/\lambda_1$). We also study the impact of this energy-efficient routing on the *route lengths* and on the *network fault tolerance*. To this end, we propose an integer linear program finding *spanners* of the topology having good stretch and two disjoint paths between all pairs of nodes. We compare the number of edges of these spanners with the one of the minimum subgraphs.

## 7.1 SNDLib Topologies

|  | $|V|$ | $|E|$ | Overprovisioning factor | | | | Tree | | Simulations | | Values given by the bounds | | | |
|---|---|---|---|---|---|---|---|---|---|---|---|---|---|---|
|  |  |  | 1 | 2 | 3 | 4 | OF | %SNE | $\lambda_1$ | $\lambda_{tree}$ | $\lambda_1$ Cut | Bound | $\delta_{max}$ | $\lambda_{tree}$ Bound |
| Atlanta | 15 | 22 | 0% | 32% | 36% | 36% | 2.66 | 36% | 38 | 101 | 3 | 38 | 4 | 88 |
| New York | 16 | 49 | 2.0% | 59% | 63% | 67% | 5.2 | 69% | 15 | 78 | 12 | 11 | 11 | 56 |
| Nobel Germany | 17 | 26 | 0% | 35% | 39% | 39% | 2.75 | 39% | 44 | 121 | 4 | 36 | 5 | 104 |
| France | 25 | 45 | 0% | 42% | 44% | 47% | 3.13 | 47% | 67 | 210 | 7 | 45 | 10 | 132 |
| Norway | 27 | 51 | 12% | 43% | 47% | 47% | 4.71 | 49% | 75 | 354 | 6 | 61 | 6 | 220 |
| Nobel EU | 28 | 41 | 12% | 32% | 34% | 34% | 2.76 | 34% | 131 | 362 | 3 | 131 | 5 | 264 |
| Cost266 | 37 | 57 | 3.5% | 32% | 35% | 37% | 3.68 | 37% | 175 | 644 | 4 | 171 | 4 | 540 |
| Giul39 | 39 | 86 | 0% | 45% | 50% | 52% | 8.25 | 56% | 85 | 702 | 11 | 70 | 8 | 340 |
| Pioro40 | 40 | 89 | 0% | 53% | 54% | 55% | 5.12 | 56% | 153 | 784 | 7 | 115 | 5 | 512 |
| Zib54 | 54 | 80 | 0% | 30% | 33% | 33% | 4.71 | 34% | 294 | 1385 | 6 | 243 | 10 | 576 |

(a) Gain of network interfaces (in %).     (b) Evaluation of the load given by the bounds.

Table 2: (*a*) Gain of network interfaces (in %) depending on the overprovisioning factor and (*b*) Evaluation of the load given by the bounds in previous section with $\lambda_1$ the capacity/demand ratio for overprovisioning factor equals 1 and $\lambda_{tree}$ for the tree.

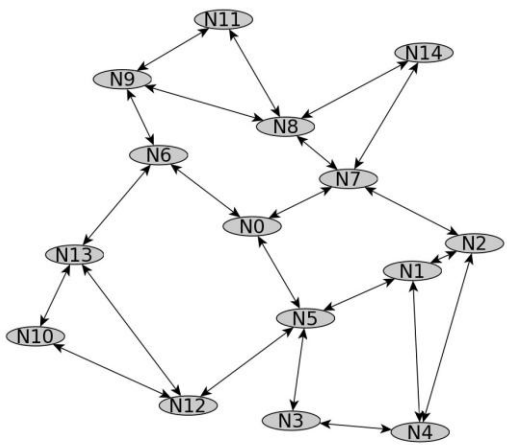

Figure 14: *Atlanta* network with 15 nodes and 22 edges.

We studied ten classical real network topologies extracted from SNDLib (http://sndlib.zib.de). These networks correspond to US (*Atlanta*, see Figure 14), European (*Nobel EU*, *Cost266*), or single country (*Nobel Germany*, *France*) topologies. Their sizes span from 15 to 54 nodes and from 22 to 89 edges, as summarized in Table 2. For these 10 topologies, we computed energy efficient routings for different capacity/demand ratios $\lambda$. We could only run the integer linear program for the smallest network, *Atlanta*, for which the results are presented in Figure 15. We see that LESS LOADED EDGE HEURISTICwell, attaining the optimal value most of the time, when the *Random* heuristic needs a larger $\lambda$ to attain the same value. As a matter of fact CPLEX already takes several hours on *Atlanta* to solve the problem for one capacity/demand ratio. We thus present the results found by the heuristic (which takes only tens of ms) in Tables 2 and 3.

**Spared network interfaces.** We give the percentage of spared network interfaces in function of the overprovisioning factor (*OF*) in Table 2 (a). A factor of 1 means that we use the minimum capacity/demand ratio necessary to route all the demands (corresponding to the value $\lambda_1$), when, e.g., a factor of 2 means that we have twice the value $\lambda_1$. Note that in most today's backbone networks overprovisioning is heavily used as it is an efficient and easy way to provide protection against failure: links are often used between 30 and 50 % of their capacity.

First, note that on some of the ten topologies, as soon as the routing is feasible (*OF*=1), some network interfaces can already be turned off (12% for *Norway* and *Nobel EU*). As a matter of fact, in this case, the important edges of the network (the edges in the minimum cut for example) are fully used, but at the same time edges at the periphery are less used and some can be spared. With an overprovisioning factor

OF=2, around one third of the edges can be spared (and even 53% for the *Pioro40* network). With larger factors (3 or 4), the gain is not as important, but still some network interfaces can be saved (e.g. 36% for *Atlanta*). We show in Table 2 (a), in the 2 last columns, the value of *OF* for which the tree is attained together with the corresponding spared network interfaces in percentage (SNE). The values are directly linked to the density of the network. For example, *Norway* needs a factor of 4.71 ($\lambda_{tree}$=4.71×$\lambda_1$) to reach the tree with 26 links (instead of 51), sparing 49% edges. Hence, the larger the density, the more the network interfaces that can be turned off. To conclude, for all the studied networks *between one third and one half of the network interfaces can be spared for usual overprovisioning factors*. Furthermore, when the best routing cannot be found by the integer linear program (large topologies), the *proposed heuristic found close to optimal solutions*.

## Full Network Topology

We report in Table 2 (b) the minimum capacity/demand ratio $\lambda_1$ for which the heuristic can perform an all-to-all routing. As explained in Section , $\lambda_1$ depends on the minimum cut of the network. We reported for each network the minimum bisection cut dividing the network into two (almost) equal-sized regions. We computed this minimum bisection cut with an integer linear program. The bound on $\lambda_1$ implied by the cut is given in the second column of the table. We see that it gives a very good indication of $\lambda_1$ for most of the networks, even if the value is not tight, as the heuristic does not always find the optimal solution. For example, we see that the bound is tight for *Atlanta*, where the minimum cut is of size 3 and splits the network into two sub-networks of sizes 8 and 7. For *Atlanta*, and *Nobel EU*, the capacity/demand ratios evaluated by the minimum cut bound is equal to the values given by the simulations ($\lambda_1$=38 for *Atlanta* and 131 for *Nobel EU*).

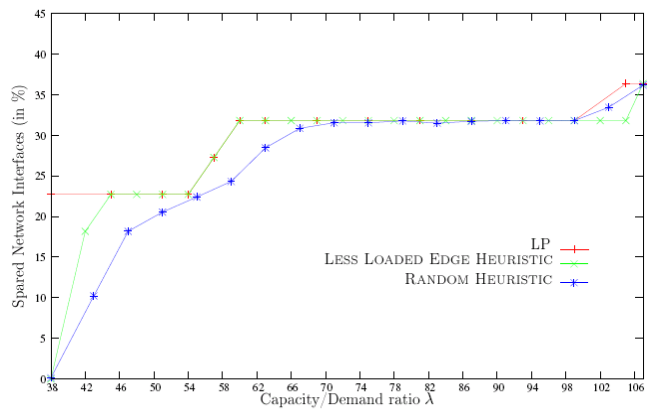

Figure 15: Percentage of spared network interfaces for the *Atlanta* network.

## Spanning Trees

As explained in Section 5, the ratio $\lambda_{tree}$, for which we get a spanning tree, depends on the network node degree. On the contrary to regular network, such as the grid, the existence of a spanning tree centered on the node of maximum degree with equal-sized branch is not given. Hence, the bounds given in Table 2 (b) are not tight. Nevertheless, we see that the node degree is still a good indication on what can be achieved on these networks. Graphs with low maximum degree attain the best configuration for lower values of capacity/demand ratio $\lambda_{tree}$. For example, the network *Atlanta* has a maximum node degree of 4 (the bound is 88) and $\lambda_{tree}$ is 101, when for *Zib54*, $\lambda_{tree}$ is 1385 for a maximum degree of 10 (the bound is 576).

|  | Overprovisioning | | | | Overprovisionning | | |
| --- | --- | --- | --- | --- | --- | --- | --- |
|  | 1 | 2 | 3 | $\frac{\lambda_{tree}}{\lambda_1}$ | 1 | 2 | 3 |
| Atlanta | 1.00 | 1.19 | 1.25 | 1.25 | 2.35 | 1.09 | 1.00 |
| New York | 1.01 | 1.24 | 1.26 | 1.32 | 4.90 | 1.24 | 1.19 |
| Nobel Germany | 1.00 | 1.11 | 1.18 | 1.18 | 2.35 | 1.04 | 1.00 |
| France | 1.00 | 1.10 | 1.12 | 1.16 | 2.48 | 1.02 | 1.01 |
| Norway | 1.02 | 1.17 | 1.18 | 1.25 | 2.61 | 1.14 | 1.04 |
| Nobel EU | 1.08 | 1.14 | 1.24 | 1.25 | 1.82 | 1.07 | 1.00 |
| Cost266 | 1.04 | 1.11 | 1.19 | 1.32 | 2.47 | 1.12 | 1.07 |
| Giul39 | 1.00 | 1.18 | 1.21 | 1.50 | 3.68 | 1.41 | 1.14 |
| Pioro40 | 1.00 | 1.25 | 1.32 | 1.42 | 4.06 | 1.12 | 1.09 |
| Zib54 | 1.00 | 1.02 | 1.07 | 1.11 | 2.16 | 1.05 | 1.01 |
| **(a) Route length** | | | | | **(b) Fault tolerance** | | |

Table 3: Impact of the energy-efficient routing on (a) the route length (Average multiplicative stretch factor) and on (b) the network fault tolerance (average number of disjoint paths).

## 7.2 Impact on the Network

We believe that network operators will implement energy efficient routing only if the impact on other parameters is limited. We discuss here the impact on the route lengths and on the fault tolerance of our proposed heuristic.

**Impact on the route lengths.** When turning-off some components in a network, we save some energy but, at the same time, we route on longer paths. The *multiplicative stretch* is defined as the ratio between the average route length in the new routing divided by the average route length with the old routing (using all the edges). Results for the SNDLib topologies are given in Table 3 (a). A stretch of one corresponds to the cases where no edge could be spared and thus the routing is not affected. We see that, as expected, the general trend is that when the overprovisioning factor increases, the paths become longer.

Nevertheless we see that the impact on the route lengths is limited. E.g., for the topology *Zib54*, the increase is 11% for the extreme case when routing on a tree (for 34% of turned-off network interfaces). In general, the increase for this extreme case spans from 11% to 50%, and in average 27%. For the network with the larger impact, *Giul39*, the stretch is increased by 50% for the tree, with a saving of 56% of the network interfaces. We see that a saving of already 45% is attained for *OF*=2, leading to a route length increase of only 18%.

**Impact on the fault tolerance.** We measure in Table 3 (b) the network fault tolerance as the *average number of disjoint paths* linking two nodes. We see that the full network topologies have an average number of disjoint paths between 1.82 and 4.90. When routing on the tree, this number is of course 1 as only one route exists, and this value is almost already attained for *OF*=3. The drop is quick as all the ten networks have a number below 1.41 for *OF*=2.

*Discussion on Technology.* When there is a failure between two nodes, it may be necessary to turn on some network interfaces to compute a new routing for some demands. Hence, this study shows that the use of such energy efficient solutions is conditioned by the existence of *technologies allowing a quick switching on of network interfaces*. Network interfaces companies are currently working on designing this kind of interfaces as mentioned in (Akyildiz, Wang, & W., 2005).

We propose in the following section a solution for the routing with fault-tolerant spanners, such that there are two disjoint paths per demand. Therefore, the impact of links failures on the network will be reduced because a protection path with enough capacity will be available.

| Network | $\alpha$ | $\beta$ | Spared Network Interfaces | OF | Mean Stretch |
|---|---|---|---|---|---|
| Atlanta | 1 | 1 | - | - | - |
| Atlanta | 1 | 2 | - | - | - |
| Atlanta | 1 | 3 | 13.64% | 1.71 | 1.36 |
| Atlanta | 2 | 1 | 13.64% | 1.74 | 1.23 |
| Atlanta | 2 | 2 | 22.73% | 1.92 | 1.42 |
| Atlanta | 2 | 3 | 22.73% | 2.21 | 1.64 |
| New York | 1 | 1 | 38.78% | 2.07 | 1.27 |
| New York | 1 | 2 | 55.10% | 3.73 | 1.75 |
| New York | 1 | 3 | 61.22% | 4.87 | 2.24 |
| New York | 2 | 1 | 55.10% | 4.27 | 2.04 |
| New York | 2 | 2 | 61.22% | 5.13 | 2.59 |
| New York | 2 | 3 | 63.27% | 6.60 | 3.17 |
| Nobel Germany | 1 | 1 | - | - | - |
| Nobel Germany | 1 | 2 | - | - | - |
| Nobel Germany | 1 | 3 | 19.23% | 2.11 | 1.44 |
| Nobel Germany | 2 | 1 | 19.23% | 2.16 | 1.39 |
| Nobel Germany | 2 | 2 | 23.08% | 2.23 | 1.67 |
| Nobel Germany | 2 | 3 | 26.92% | 2.39 | 1.73 |
| France | 1 | 1 | - | - | - |
| France | 1 | 2 | 17.78% | 2.30 | 1.25 |
| France | 1 | 3 | 24.44% | 2.63 | 1.55 |
| France | 2 | 1 | 24.44% | 2.51 | 1.62 |
| France | 2 | 2 | 28.89% | 3.06 | 1.84 |
| France | 2 | 3 | 33.33% | 3.27 | 2.02 |

Table 4: ($\alpha,\beta$)−spanners with 2 disjoint paths between every pair of nodes.

## 7.3 Fault tolerant spanners

During the simulations, we show that as soon as some network interfaces are spared, the average number of disjoint paths fail dramatically. Therefore, the network fault tolerance is no longer assured, as any failure will imply the re-computation of the routing structure. As the network tolerance is of major importance, we show in this subsection how to find a fault tolerant spanner which has at least $\gamma$ disjoint paths for each demand and which respects additive and multiplicative constraints for the stretch.

We show in the following that this structure is related to the ($\alpha,\beta$)−spanners (D. Peleg and A.A. Schaffer. (1989)). An ($\alpha,\beta$)−spanner of a graph $G=(V,E)$ is a subgraph $H=(V(H) \subseteq V, E(H) \subseteq E)$ of $G$ such that $\forall (u,v) \in V \times V$, $d_H(u,v) \leq \alpha \cdot d_G(u,v) + \beta$ in which $|E(H)|$ is minimized. This guarantees that the distances of any couple of vertices in the spanner is streched by a multiplicative and additive factors α and β, ideally close to 1 and 0.

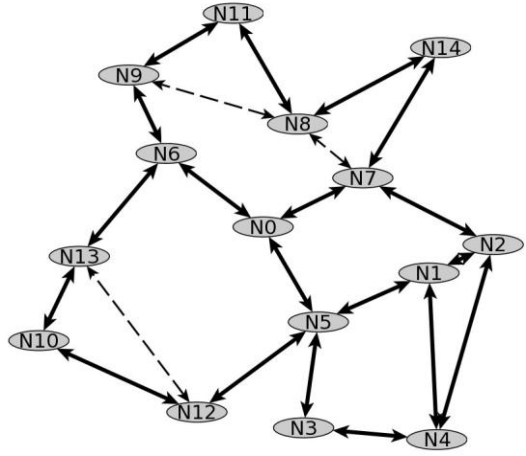

Figure 16: A (1,3)−spanner for *Atlanta* with 2 disjoint paths for each demand represented in full lines.

This problem can be modeled as previously seen, with a multicommodity *integral flow problem* in which the objective function is the minimization of the number of edges. The demands $\mathcal{D}_{st}$ are equal to γ between every pair of vertices $(s,t) \in V \times V$. As the utilization constraint imposes only unitary flow per edge and per demand, γ disjoint paths will be set. Finally, the distance constraint limits the maximum route length that can be accepted per demand. Note that this constraint is given in a global manner for all the γ disjoint paths.

The integer linear program can be modeled as follows.

The *Objective function* is

$$\min \sum_{e \in E} x_e$$

subject to:

Flow constraints: $\forall (s,t) \in V \times V, \forall u \in V,$

$$\sum_{v \in N(u)} f_{vu}^{st} - \sum_{v \in N(u)} f_{uv}^{st} = \begin{cases} -\gamma & \text{if} \quad u = s \\ \gamma & \text{if} \quad u = t \\ 0 & \text{otherwise} \end{cases}.$$

Utilization constraint: $\forall (s,t) \in V \times V, \forall e = (u,v) \in E,$

$$f_{uv}^{st} + f_{vu}^{st} \leq x_e$$

Distance constraint : $\forall (s,t) \in V \times V$,

$$\sum_{(u,v) \in E} f_{uv}^{st} \leq \gamma(\alpha d_G(s,t) + \beta).$$

Figure 16 shows a (1,3)−spanner with 2 disjoint paths (γ=2) between any pair of vertices. Consider for example the distance $d_H$(N5,N13) between N5 and N13. In G, $d_G$(N5,N13)=2, and in the spanner H, the two disjoint paths, (N5,N0,N6,N13) and (N5,N12,N10,N13) have both distances 3. This respects the distance constraint:

$$\sum_{(u,v) \in E(H)} f_{uv}^{N5N13} = 6 \leq 2 \times (1 \times d_G(N5,N13) + 3) = 10.$$

This spanner provides a gain of 13.64% of the number of network interfaces.

Table 4 shows the results of the integer linear program for *Atlanta*, *New York*, *Nobel Germany* and *France*. The number of disjoint paths per demand γ=2. As the program takes several hours to be launched on large topologies, we set a limit of 1 hour, and a gap from the optimal solution of 3% for the integer linear program. The results show that for *Atlanta*, 23% of network interfaces can still be spared. The spanner needs an overprovisionning factor of 1.92 and an average stretch of 1.42 is attained. For *Atlanta*, the integer linear program was not able to find (1,1), (1,2) and (2,0) fault tolerant spanners. This is mainly due to topology constraints where 2 disjoint paths with such distance constraints cannot be found.

Note that we compute the average stretch by the comparison with fault-tolerant spanners with two disjoint paths per demand, without the distance constraint, and with the objective of minimization of $\sum_{\substack{(s,t) \in V \times V, \\ (u,v) \in E}} f_{uv}^{st}$.

To conclude this section, fault-tolerant spanners are interesting solutions to achieve power consumption reduction while taking into account network fault tolerance and distance constraints.

# 8 Conclusion and Perspectives

In this work, we present through a simplified architecture the problem of minimizing power consumption in networks. We show non-approximation results. The simulations on real topologies show that the gain in energy is significant when some network interfaces can be turned-off.

- *At least one third of the network interfaces can be spared for usual range of demands.*

- For a medium-sized backbone network, this leads to a *reduction of power consumption* of approximately 33MWh per year (for *Cost266* with 37% of spared interfaces). For this estimation, we consider a scenario where the turned-off interfaces of our simplified architecture are 4-port Gb Ethernet linecards. We believe it corresponds to a reasonable capacity for backbone networks. We use the consumption values given in (Chabarek, et al., 2008): 100W for these linecards.

- The *route lengths* increase, but not too much: in average 27% for almost all studied topologies.

- *Fault tolerance* can be achieved with the use of *fast switching-on technologies* or by adding *disjoint path constraints* to the problem so that the network remains γ-connected, allowing a tolerance of γ−1 failures.

- The bounds for specific topologies are useful for general networks to evaluate the overprovisioning factors needed for such energy-efficient solutions.

As part of future work, we plan to study a detailed cost function for a more complex router architecture. Moreover, we will carry on lab experiments on small network topologies to measure in practice the performance of the proposed energy-efficient routing.

# Acknowledgements

This work has been partly funded by ANR DIMAGREEN[1] (DesIgn and MAnagement of GREEN networks with low power consumption).

---

[1] http://www-sop.inria.fr/teams/mascotte/Contrats/DIMAGREEN/wiki/

# Key Terms and Definitions

**APX:** in complexity theory, the class APX ("approximable") is the set of optimization problems that admit a polynomial-time approximation algorithms with approximation ratio bounded by a constant. In other words, if a problem is not in APX, it means that there does nos exist a polynomial-time within a constant factor of approximation (unless P=NP).

**Heuristic:** algorithm that finds quickly a solution to a problem. For NP-hard problems, heuristic algorithms are used to find in polynomial time a solution, not necessaily optimal, but a solution that may be not so bad.

**ICT:** Information and Communication Technology is a term that includes any communication device or application: radio, television, cellular phones, computer and network hardware and software, satellite systems, and so on.

**Complete graph:** in graph theory, a complete graph is an undirected graph with all possible edges. In other words, if any two nodes of the graph are linked by an edge, then the graph is a complete graph.

**Grid:** in graph theory, a grid graph is the square product of two paths.

**Linear Programming:** mathematical method for describing an optimization problem in a mathematical model. Linear Programming is a technique for optimization problems with linear objective functions, subject to linear equalities and linear inequalities.

**Multicommodity flow problem:** network flow problem with multiple commodities (demands) between different sources and sinks. In graph theory, a flow network is a directed graph where each edge has a capacity and each edge receives a flow. This amount of flow cannot exceed the capacity of the edge. Given a graph, a set of demands, and the capacities of the edges, the mulicommodity flow problem consists in finding a flow network satisfying the demands and respecting the constraints of capacities.

**Network:** set of nodes linked by links. A communication network is a set of nodes connected by links which permit telecommunication between the nodes  A network can be seen as a graph composed by a set of vertices and a set of edges.

**NP-Complete problem:** in complexity theory, the complexity class NP-complete is a class of decision problems. A decision problem is NP-complete if it is in NP (class of problems such that any given solution can be verified in polynomial time) and also it is in the set of NP-hard problems.

**Routing:** set of paths of the network (graph) in which traffic will be sent.

**Tree:** in graph theory, a tree is an undirected graph in which two vertices are connected by exactly one simple path. In other words, any connected graph without cycles is a tree.

## Additional Reading